\documentclass[prl,twocolumn,showpacs,preprintnumbers,amsmath,amssymb]{revtex4-1} 
\usepackage{graphicx}
\usepackage{dcolumn}
\usepackage{bm}
\newcommand{\xx}{\mathbf{x}}
\newcommand{\kk}{\mathbf{k}}
\begin{document}
\preprint{TCC-040-10}
\title{Non-Gaussianity Consistency Relation for Multi-field Inflation}
\author{Naonori Suma Sugiyama}
\email{sugiyama@astr.tohoku.ac.jp}
\affiliation{Astronomical Institute, Graduate School of Science, Tohoku
University, Sendai 980-8578, Japan}  

\author{Eiichiro Komatsu}
\affiliation{Texas Cosmology Center and the Department of Astronomy,
The University of Texas at Austin, 1 University Station, C1400, Austin,
TX 78712}

\author{Toshifumi Futamase}
\affiliation{ Astronomical Institute, Graduate School of Science, Tohoku
University, Sendai 980-8578, Japan}
%\date{\today}
\begin{abstract}
 While detection of the ``local form'' bispectrum of primordial perturbations would rule out all single-field inflation models, multi-field models would still be allowed. We show that multi-field models described by the $\delta N$ formalism obey an inequality between $f_{\rm NL}$ and one of the local-form {\it trispectrum} amplitudes, $\tau_{\rm NL}$, such that $\tau_{\rm NL}>\frac12(\frac65f_{\rm NL})^2$ with a possible logarithmic scale dependence, provided that 2-loop terms are small. Detection of a violation of this inequality would rule out most of multi-field models, challenging inflation as a mechanism for generating the primoridal perturbations.
\end{abstract}

\pacs{Valid PACS appear here}% PACS, the Physics and Astronomy
\maketitle
Can we rule out inflation as a mechanism for generating primordial
curvature perturbations? 
Inflation is
indispensable for explaining 
homogeneity and flatness of the observable universe
\cite{guth:1980,*linde:1982,*linde:1983,*albrecht/steinhardt:1982}. Yet,
its predictions for the statistical properties of primordial curvature
perturbations may be falsifiable. 

The basic predictions that inflation generates 
adiabatic, nearly scale-invariant, and nearly Gaussian primordial
curvature perturbations 
\cite{mukhanov/chibisov:1981,*guth/pi:1982,*hawking:1982,*bardeen/turner/steinhardt:1983,starobinsky:1982} are all consistent with the current observations
(see, e.g., \cite{komatsu/etal:2009,*komatsu/etal:prep}). 
Notably, many inflation models predict that the amplitude of
fluctuations on large scales is greater than that on small scales. In
terms of the power spectrum of primordial curvature perturbations
$\zeta$, we say $k^3P_\zeta(k)\propto k^{n_s-1}$ with $n_s<1$. 
The power spectrum is defined by $\langle
\zeta_{\kk}\zeta_{\kk'}\rangle=(2\pi)^3\delta(\kk+\kk')P_\zeta(k)$. 
The
latest observations give $n_s= 0.96\pm 0.01$
\cite{komatsu/etal:prep,dunkley/etal:prep}, which may be taken as
evidence for inflation.  

The future, more sensitive experiments will continue to test the other
predictions: adiabaticity and Gaussianity of fluctuations. In this
paper, we shall focus on the latter.
Departure from Gaussianity, called {\it non-Gaussianity}, has emerged as
a powerful test of inflation over the last decade (see
\cite{komatsu:phd,*bartolo/etal:2004,*komatsu/etal:astro2010,*koyama:2010,*chen:2010,*wands:2010} for reviews).

One of the major theoretical discoveries made from these studies is
that {\it all} single-field inflation models yield a specific amount of
non-Gaussianity in the so-called squeezed limit of the bispectrum
(Fourier transform of the three-point correlation function) of $\zeta$,
given by $f_{\rm NL}=\frac5{12}(1-n_s)\simeq 0.02$
\cite{creminelli/zaldarriaga:2004} (also see \cite{maldacena:2003,*acquaviva/etal:2002,*seery/lidsey:2005,*chen/etal:2007,*cheung/etal:2008,*ganc/komatsu:2010,*renaux-petel:2010}). Here, $f_{\rm NL}$ 
characterizes the amplitude of the so-called ``local form'' bispectrum
\cite{gangui/etal:1994,*verde/etal:2000,komatsu/spergel:2001}:
\begin{equation}
 B_\zeta=\frac65f_{\rm
  NL}\left[P_\zeta(k_1)P_\zeta(k_2)+(\mbox{2 perm}.)\right],
\end{equation} 
where 
$\langle \prod_{i=1}^{3}\zeta(\kk_i)\rangle 
   = (2\pi)^3 \delta^{3}(\sum_i\kk_i) B_\zeta(k_1,k_2,k_3)$, and
   the ``squeezed limit'' is given by taking $k_3\ll k_1\approx k_2$,
   i.e., $B_\zeta(k_1,k_2,k_3)\to \frac{12}5f_{\rm
   NL}P_\zeta(k_1)P_\zeta(k_3)$. All single-field inflation models
   predict $(1-n_s)P_\zeta(k_1)P_\zeta(k_3)$ in this limit.

The current best limit is $f_{\rm NL}=32\pm 21$ (68\%~CL;
\cite{komatsu/etal:prep}). 
As various second-order effects generate $f_{\rm NL}={\cal O}(1)$ (see
\cite{komatsu:2010} for a review and references therein), 
a convincing detection of $f_{\rm NL}\gg 1$ would rule out all
single-field inflation models. The Planck satellite is expected to
reduce the error bar by a factor of four \cite{komatsu/spergel:2001}.

However, detection of $f_{\rm NL}$ would not rule out {\it
multi}-field models. How can we test them also?  
Our work in this paper is motivated by the Suyama-Yamaguchi inequality,
$\tau_{\rm 
NL}\ge (\frac65f_{\rm NL})^2$ \cite{suyama/yamaguchi:2008}. Here,
$\tau_{\rm NL}$ is one of the amplitudes of the local-form {\it
trispectrum} defined by \cite{boubekeur/lyth:2005}
\begin{equation}
  T_{\zeta} = \tau_{\rm NL}[P_{\zeta}(|\kk_1 + \kk_3|) P_{\zeta}(k_3)
			P_{\zeta}(k_4) + (\mbox{11 perm.})],
\end{equation}
where 
$\langle \prod_{i=1}^{4}\zeta(\kk_i)\rangle 
   =(2\pi)^3 \delta^{3}(\sum_i\kk_i) T_\zeta(k_1,k_2,k_3,k_4)$.

As emphasized in \cite{komatsu:2010}, if the new experimental data (such
as Planck) detect $f_{\rm NL}$ (hence ruling out single-field models)
but do not see $\tau_{\rm NL}$ large enough to satisfy the above
inequality, then a large class of multi-field models may be ruled out.
The crucial question is then, ``how generic is the Suyama-Yamaguchi
inequality?'' It was pointed out in \cite{komatsu:2010} that this
inequality may not be generic enough, as there are cases where this
inequality is not satisfied.  Recently, Suyama et al. \cite{suyama/etal:prep}
considered the same issue, where they have truncated the $\delta N$ expansion
(given below) at the second order and have considered a part of 1-loop
corrections. 
The goal of this paper is to find a more general inequality than theirs.
We shall retain the terms up to the fourth order of $\delta N$ expansion, as
these terms are required for the consistent calculations up to the
1-loop level. As a result, we find a weaker bound than the original
Suyama-Yamaguchi inequality. 
This is relevant because, as shown in
\cite{Heiner/etal:2008,*Yeinzon/etal:2010}, large and observable
primordial non-Gaussianity can be generated when the loop contributions
dominate over the tree contributions in the bispectrum and/or in the
trispectrum. 

Throughout this paper, we shall consider a class of multi-field models
which satisfy the following conditions:
\begin{itemize}
 \item[1.] Scalar fields are responsible for generating curvature
	   perturbations; thus, potential contributions from vector
	   fields (see \cite{dimastrogiovanni/etal:2010} for a review
	   and references therein) are ignored.
 \item[2.] Fluctuations in scalar fields at the horizon crossing are
	   scale invariant and Gaussian.
\end{itemize}
Therefore, we assume that non-Gaussianity is generated only on super
horizon scales, 
according to the $\delta N$ formalism \cite{starobinsky:1982,starobinsky:1986,*salopek/bond:1990,*sasaki/stewart:1996,*lyth/rodriguez:2005}.
While the ``quasi-single-field inflation'' model proposed by Chen and Wang
\cite{chen/wang:2010,*chen/wang:2010b} is an example to which this
condition may not apply, their model yields $\tau_{\rm NL}\gg f_{\rm
NL}^2$, satisfying the inequality.
Yet, the condition 2 is probably too strong. Whether this condition can be 
relaxed significantly merits further investigation.

According to the $\delta N$ formalism, the curvature perturbation,
$\zeta$, is given by derivatives of the number of $e$-fold,
$N(t,t_*)=\int_{t_*}^t Hdt'$, with respect to scalar fields,
$\varphi^a$, at the
horizon-crossing time $t_*$ ($a_{\ast}H_{\ast} = k$):
\begin{equation}
  \zeta(\xx,t) = N_a(t,t_{\ast}) \delta \varphi^a_{\ast}(\xx) +
   \frac{1}{2}N_{ab}(t,t_{\ast}) \delta \varphi^a_{\ast}(\xx) \delta
   \varphi^b_{\ast}(\xx) \cdots
\label{eq:deltaN}
\end{equation}
where $\delta \varphi_{\ast}^a$ is  a fluctuation of $\varphi^a$ evaluated at
$t_{\ast}$, i.e., $\delta \varphi_{\ast}^a(\xx) \equiv \delta \varphi^a
(t_{\ast},\xx)$. 
Note that $N_a\equiv \partial N/\partial\varphi^a_{\ast}$ and 
$N_{ab}\equiv \partial^2 N/\partial\varphi^a_{\ast}\partial\varphi^b_{\ast}$.

The second condition above implies that the power spectrum of scalar
fields is given by
\begin{equation}
  \langle \delta \varphi^a_{\kk}(t_{\ast}) \delta
   \varphi^b_{\kk^{\prime}}(t_{\ast}) \rangle = (2\pi)^3 \delta^{(3)}
   (\kk + \kk^{\prime}) \delta^{ab}
\frac{2\pi^2}{k^3}\mathcal{P}_{\ast},
\label{eq:Pphi}
\end{equation}
where $\mathcal{P}_{\ast}\equiv (H_*/2\pi)^2$.
Note that we have assumed that scalar field fluctuations with different
indices are uncorrelated, $\langle \delta \varphi^a \delta
   \varphi^b \rangle\propto\delta^{ab}$. This can
   be done without loss of generality: we could, for example, write the
   correlation matrix as $\langle \delta \varphi^a \delta
   \varphi^b \rangle\propto M^{ab}$, where $M$ is a real positive symmetric
   matrix. One can then diagonalize $M$ as $M=UDU^t$. Redefining scalar
   field fluctuations as $\delta\varphi\to
   \tilde{\delta\varphi}=U\delta\varphi$ will recover
   Eq.~(\ref{eq:Pphi}).

Now, we impose the third condition:
\begin{itemize}
 \item[3.] Truncate the $\delta N$ expansion [Eq.~(\ref{eq:deltaN})] at
	   the order of $\delta\varphi^4$, 
	   i.e., $\zeta = N_a \delta \varphi^a_{\ast} + \frac{1}{2}N_{ab} \delta \varphi^a_{\ast} \delta \varphi^b_{\ast} 
+ \frac{1}{3!} N_{abc} \delta \varphi^a_{\ast} \delta \varphi^b_{\ast} \delta \varphi^c_{\ast}
+ \frac{1}{4!} N_{abcd} \delta \varphi^a_{\ast} \delta \varphi^b_{\ast}
	   \delta \varphi^c_{\ast} \delta \varphi^d_{\ast} .$
Thus, we shall ignore the contributions in the power spectrum, bispectrum, or trispectrum coming from $O(\delta\varphi^5)$. 
\end{itemize}
The 4th-order term is needed when we calculate all of the
1-loop contributions in $f_{\rm NL}$ and $\tau_{\rm NL}$.
In the following, we shall include all of the 1-loop contributions,
while some of the higher-order loop contributions are also included.

The power spectrum is given, up to the 4th order, by
\begin{align}
\mathcal{P}_{\zeta} = & \mathcal{P}_{\ast}\Big[  N_aN_a  + {\rm Tr}(N^2) \mathcal{P}_{\ast} \ln(k L) 
+ N_aN_{abb}\mathcal{P}_{\ast} \ln(k_{\rm max} L) \notag \\
& \ \ \ \  + \frac{1}{4}N_{acc}N_{abb}\mathcal{P}^2_{\ast} \ln^2(k_{\rm max} L)  \notag \notag \\
& \ \ \ \  +N_{abcc}N_{ab} \mathcal{P}^2_{\ast} \ln(k L) \ln(k_{\rm max} L) \dots \Big],
\label{power}
\end{align}
where we have used the following notations: $N_aN_a\equiv \sum_a N_a^2$ and ${\rm Tr}(N^2) \equiv \sum_{ab} N_{ab} N_{ab}$.
The $L$ is a finite size of a box which is chosen to be much larger than
the region of interest, such that the condition $L k \gg 1$ is satisfied
for arbitrary $k$, and $k_{\rm max}$ is the ultra-violet cutoff.
The 1st term is the tree contribution; the 2nd and 3rd terms
are the 1-loop contributions; and the 4th and 5th terms are the 2-loop
contributions. 

This result can be simplified by using the following quantities 
(see Eq.~(25) of \cite{byrnes/etal:2007}):
\begin{align}
&\tilde{N}_a \equiv N_a + \frac{1}{2}N_{abb} \mathcal{P}_{\ast}
 \ln(k_{\rm max} L), \\ 
&\tilde{N}_{ab} \equiv N_{ab} + \frac{1}{2}N_{abcc} \mathcal{P}_{\ast}
 \ln(k_{\rm max} L).
\end{align}
Then Eq.~(\ref{power}) becomes
\begin{equation}
\label{eq:pzeta}
\mathcal{P}_{\zeta} = \tilde{N}_a \tilde{N}_a \mathcal{P}_{\ast}  \left( 1 + \mathcal{P}_{\rm loop} + \dots \right),
\end{equation}
where we have defined a positive-definite quantity
\begin{equation}
\mathcal{P}_{\rm loop} \equiv \frac{ {\rm Tr} (\tilde{N}^2)}{\tilde{N}_a \tilde{N}_a} \mathcal{P}_{\ast} \ln(k L).
\end{equation}
Here, the dots in Eq.~(\ref{eq:pzeta}) include the higher-order terms
such as $N_{abcc}^2{\cal P_*}^2$.
This is a nice way of writing the power spectrum etc., as the results
do not include the ultra-violet cutoff, $k_{\rm max}$, explicitly: the
cutoff can be absorbed by redefining the derivatives of $N$.

As we can take $L$ such that $kL\gg 1$, $\mathcal{P}_{\rm loop}$ is
essentially a constant factor, rescaling the overall amplitude of the
power spectrum without destroying the observed scale invariance of the power
spectrum. Without loss of generality, we shall take $k$ to
be the usual normalization scale used by the {\sl WMAP} collaboration,
$k_0=0.002~{\rm Mpc}^{-1}$.

Kawakami et al. \cite{kawakami/etal:2009} have derived the expressions
for $f_{\rm NL}$ and $\tau_{\rm NL}$ up to the 4th order
(also see \cite{suyama/etal:prep}).
These expressions are again simplified by using the redefinition of the
derivatives of $N$ and ignoring the higher-order terms:
\begin{align}
\frac{6}{5}f_{\rm NL} &\simeq  \left[ \tilde{N}_a\tilde{N}_a +  {\rm Tr} (\tilde{N}^2) \mathcal{P}_{\ast}\ln(k_0 L)  \right]^{-2} \notag \\
& \times \Big[ \tilde{N}_a \tilde{N}_b \tilde{N}_{ab} + \left( {\rm Tr} (\tilde{N}^3) + 2\tilde{N}_a \tilde{N}_{bc} \tilde{N}_{abc} \right) \mathcal{P}_{\ast} \ln(k_0 L) \Big], 
\label{fNL}
\end{align}
\begin{align}
 \tau_{\rm NL} &\simeq  \left[  \tilde{N}_a\tilde{N}_a +  {\rm Tr} (\tilde{N}^2) \mathcal{P}_{\ast}\ln(k_0 L)  \right]^{-3}\notag \notag \\
             & \times \Big[\tilde{N}_a \tilde{N}_{ab}\tilde{N}_{bc} \tilde{N}_c 
                   +  \Big( 2\tilde{N}_a \tilde{N}_{ab} \tilde{N}_{cd} \tilde{N}_{bcd}  +  {\rm Tr} (\tilde{N}^4)  \notag \\
             &  + 2\tilde{N}_a \tilde{N}_{bc} \tilde{N}_{bd} \tilde{N}_{acd}  
               + \tilde{N}_{a} \tilde{N}_b \tilde{N}_{acd} \tilde{N}_{bcd}  \Big) \mathcal{P}_{\ast} \ln(k_{0} L)  \Big],
\label{tNL}
\end{align}
where $\tilde{N}_{abc} \equiv N_{abc} + \frac{1}{2}N_{abcdd} \mathcal{P}_{\ast}
 \ln(k_{\rm max} L)$.
Although the loop terms of the bispectrum and trispectrum have
terms like $\ln(k_b L)$, $\ln(k_t L)$ and $\ln(k_p L)$
where $k_b \equiv {\rm min} \left\{ k_i \right\}$ with $i = \{1,2,3\}$ or $\{1,2,3,4\}$ 
, $k_t \equiv{\rm min}  \{k_i,|\vec{k}_j + \vec{k}_l|  \}$ with
$(i,j,l) = \{1,2,3,4  \}$ and $\ln(k_p L) \sim \ln(k_i L) \sim \ln(|\vec{k}_j + \vec{k}_l| L)$ with $(i,j,l) = \{1,2,3,4  \}$, 
we assume that these are similar to
$\ln(k_0 L)$, i.e., $\ln(k_0 L)\sim \ln(k_b L)\sim \ln(k_t L) \sim \ln(k_p L)$.
From now on, we shall remove the tildes from the equations, i.e.,
$\tilde{N}\rightarrow N$. 

Now, we are ready to derive the new inequality. 
First of all, we use the inequality between arbitrary real numbers
$\alpha$ and $\beta$: $ \alpha^2 + \beta^2 \geq \frac{1}{2}\left( \alpha
+ \beta \right)^2$. Choosing $\alpha$ and $\beta$ as 
\begin{align}
&\alpha \equiv \left[ N_aN_a(1 + \mathcal{P}_{\rm loop}) \right]^{-2} \left[ N_a N_b N_{ab} + N_aN_{bc}N_{abc}\mathcal{P}_{\ast}\ln(k_0 L) \right], \notag \\
&\beta  \equiv \left[ N_aN_a(1 + \mathcal{P}_{\rm loop}) \right]^{-2}\left[ {\rm Tr} (N^3) + N_aN_{bc}N_{abc} \right] \mathcal{P}_{\ast} \ln(k_0 L),
\end{align}
we find
\begin{align}
&\left[ N_aN^a(1+{\cal P}_{\rm loop})\right]^{-4} \notag \\
& \times \Big[ \Big( N_a N_b N_{ab} + N_a N_{bc} N_{abc}
 \mathcal{P}_{\ast} \ln(k_0 L)\Big)^2   \label{0th} \\
& \ \  + \left( {\rm Tr} (N^3) + N_a N_{bc} N_{abc} \right)^2 \mathcal{P}^2_{\ast} \ln^2(k_0 L) \Big] 
 \geq \frac{1}{2}\left( \frac{6}{5}f_{\rm NL} \right)^2.
\notag
\end{align}

Next, pick up the first term of the LHS in (\ref{0th}), and use the
Cauchy-Schwarz inequality. When we define the inner product of arbitrary
vectors $v_a$ and $u_b$ as $\langle v,u \rangle \equiv \sum_a v_a u_a$, 
then the Cauchy-Schwarz inequality leads to $\langle v,u \rangle^2 \leq \langle v,v \rangle \langle
u,u \rangle$. Choosing $v_a$ and $u_a$ as
$v_a \equiv N_a$ and $u_a \equiv N_b N_{ba} +  N_{bc}
N_{abc}\mathcal{P}_{\ast} \ln(k_{0}L )$, we find
\begin{align}
& \frac{ \Big(   N_{a}N_b N_{ab} + N_aN_{bc} N_{abc}\mathcal{P}_{\ast} \ln(k_{0}L )  \Big)^2}{(N_aN_a)^4  (1 + \mathcal{P}_{\rm loop})^4}  \notag\\
& < \frac{N_b N_{ba} N_{ad}N_d +   2N_dN_{da}N_{abc}N_{bc} \mathcal{P}_{\ast} \ln(k_{0} L) 
}{  (N_aN_a)^3 \left( 1 + \mathcal{P}_{\rm loop} \right)^3}  \notag  \\
& \ \ \ \ \  +\frac{ N_{ab}N_{abc}N_{cde}N_{de}\mathcal{P}_{\ast}^2
 \ln^2(k_0 L)}{(N_aN_a)^3 (1 + \mathcal{P}_{\rm loop})^3},
\label{1st}
\end{align}
where we have also used $1/(1 + \mathcal{P}_{\rm loop}) < 1$ with
$\mathcal{P}_{\rm loop} > 0$ on the RHS. Note that the last term on
the RHS is a 2-loop contribution, which becomes important later.

Finally, pick up the second term of the LHS in (\ref{0th}), and use the
Cauchy-Schwarz inequality again: for arbitrary real symmetric matrices
$M$, $L$, we have ${\rm  Tr}^2(LM)\leq {\rm Tr} (M^2) {\rm Tr }(L^2)$.
Choosing $L$ and $M$ as $L_{ab} \equiv N_{ab}$ and $M_{ab} \equiv
N_{ac}N_{cb} + N_{c}N_{cab} $, we find
\begin{align}
&\frac{ \left( {\rm Tr} (N^3) + N_aN_{bc} N_{abc} \right)^2 \mathcal{P}^2_{\ast} \ln^2(k_{0}L ) }
{(N_aN_a)^4 (1 + \mathcal{P}_{\rm loop})^4}  \notag \\
& <  \frac{  \left( {\rm Tr}(N^4) + 2N_{ac}N_{cb}N_{dab}N_d + N_{c}N_{cab}N_{abd}N_d \right)\mathcal{P}_{\ast}\ln(k_0 L) }
{(N_aN_a)^3 (1 + \mathcal{P}_{\rm loop})^3},
\label{2st}
\end{align}
where we have also used $\mathcal{P}_{\rm loop}/(1 + \mathcal{P}_{\rm
loop}) < 1 $.
Here, let us reconsider the effect of our approximation that all the  
logarithmic
factors are similar: $\ln(k_0L)\sim  \ln(k_bL)\sim  \ln(k_tL)\sim   
\ln(k_pL)$.
If we relax this assumption, then we should replace $ \ln(k_0L)$ in
the right hand side of Eq. (\ref{2st}) with $ \ln(k_0L)\to \ln(k_tL)R 
$, where
$R \equiv \ln^2(k_b L)/\ln(k_t L) \ln(k_p L)$. Therefore, our  
approximation
is valid also when the geometric mean of $\ln(k_t L)$ and  $\ln(k_p L) 
$ is
similar to $\ln(k_b L)$ (but not necessarily $\ln(k_t L)\sim \ln(k_p L) 
$).

Collecting these results, we obtain
\begin{equation}
 \tau_{\rm NL} + (2~{\rm loop}) > \frac{1}{2} \left( \frac{6}{5}f_{\rm
					       NL} \right)^2,
\end{equation}
where the ``2~loop'' term is the last term in the RHS of
Eq.~(\ref{1st}).
This result shows that, when we allow ourselves for completely general
models in which this particular 2-loop term can become important, the
Suyama-Yamaguchi inequality, 
$\tau_{\rm NL}\ge  \left( \frac{6}{5}f_{\rm NL}
\right)^2$, may be violated badly. This illustrates the limitation of
this inequality. 

Still, from a model-building point of view, it is reasonable to assume
that the 2-loop terms are sub-dominant compared to the tree or 1-loop
terms; otherwise, we would have to require fine-tunings between the
derivatives of $N$.
Let us then study the consequence of ignoring this particular
2-loop term. We shall impose the following conditions:
\begin{align}
&\frac{N_{ab}N_{abc}N_{cde}N_{de}{\cal P}_{\ast}^2 \ln^2(k_0L)}{N_{b}N_{ba}N_{ac}N_c} \ll 1, \notag \\
&\frac{N_{ab}N_{abc}N_{cde}N_{de}{\cal P}_{\ast}^2 \ln^2(k_0L)}{N_{b}N_{ba}N_{ac}N_c} \ll  \Bigg| \frac{N_dN_{da}N_{abc}N_{bc}{\cal P}_{\ast} \ln(k_0 L)}{N_bN_{ba}N_{ac}N_c}\Bigg|.
\label{con}
\end{align}
The first condition is (tree)$\gg$(2-loop), and the second is
(1-loop)$\gg$(2-loop) for the terms in the RHS of Eq.~(\ref{1st}).
Interestingly, from the Cauchy-Schwarz inequality for $N_{ab}N_b$ and
$N_{abc}N_{bc}$, we find
\begin{align}
 &\Bigg( \frac{N_dN_{da}N_{abc}N_{bc}{\cal P}_{\ast} \ln(k_0 L)}{N_bN_{ba}N_{ac}N_c} \Bigg)^2 \notag \\
 & \ \ \leq \frac{N_{ab}N_{abc}N_{cde}N_{de}{\cal P}_{\ast}^2 \ln^2(k_0L)}{N_{b}N_{ba}N_{ac}N_c} \notag \\
 & \ \ \ll  \Bigg| \frac{N_dN_{da}N_{abc}N_{bc}{\cal P}_{\ast} \ln(k_0 L)}{N_bN_{ba}N_{ac}N_c} \Bigg|,
 \label{rel}
\end{align}
from which we obtain the following bound on a particular form of 1-loop
contributions: 
\begin{equation}
\Bigg| \frac{ N_dN_{da}N_{abc}N_{bc}{\cal P}_{\ast} \ln(k_0 L) }{N_bN_{ba}N_{ac}N_c} \Bigg| \ll 1.
\label{aho}
\end{equation}
As a result, if we ignore the last term in the RHS of Eq.~(\ref{1st}),
we must also ignore the second term, leaving only the tree-level term in
the RHS of Eq.~(\ref{1st}). This is a peculiar feature of these terms,
whose physical meaning is not clear. 

In any case, provided that the following additional condition is met:
\begin{itemize}
\item [4.] The 2-loop contributions are sub-dominant compared to the
      tree-level or 1-loop contributions (or at least the particular
      2-loop term in 
      the RHS of Eq.~(\ref{1st}) is small compared to the others),
\end{itemize}
we finally arrive at the new inequality:
\begin{equation}
  \tau_{\rm NL} > \frac{1}{2} \left( \frac{6}{5}f_{\rm NL} \right)^2,
  \label{main}
\end{equation}
which is the main result of this paper, and is valid as long as the
2-loop contributions are small.
This result generalizes the
Suyama-Yamaguchi inequality (which included only the tree-level terms)
as well as Ref.~\cite{suyama/etal:prep} (which included up to the
second-order terms). This relation can have a logarithmic
scale dependence via $R=\ln^2(k_b L)/[\ln(k_t L) \ln(k_p L)]$.

What are the implications for inflation? In principle, if the following
relation is observed, 
\begin{equation}
  \frac{1}{2} <  \frac{\tau_{NL}}{\left( \frac{6}{5}f_{NL} \right)^2} \leq 1,
\end{equation}
then it implies that there was a non-negligible contribution from 1-loop
terms, which would help constrain the physics of multi-field models
(via the forms of $N$). However, the most interesting case would be the
observation of a complete violation of the inequality, i.e., 
\begin{equation}
   \tau_{NL} \ll \frac{1}{2}\left( \frac{6}{5}f_{NL} \right)^2,
\end{equation}
which implies that 
inflation cannot be responsible for generating the observed
fluctuations, {\it provided that} (1) scalar fields are the source of
fluctuations; (2) fluctuations at the horizon crossing are scale
invariant and Gaussian; (3) the evolution of fluctuations obeys the
$\delta N$ formalism; and (4) the 2-loop contributions are small.

We may not be so far away from testing this prediction. If the value of
$f_{\rm NL}$ is as large as what is implied from the current data,
$f_{\rm NL}\sim 30$, then the threshold value, $\tau_{\rm NL}\sim 650$, is close
to the 2-$\sigma$ limit expected from {\it Planck}
\cite{kogo/komatsu:2006,smidt/etal:2010}. The large-scale structure
observations should also help improving the limits on $\tau_{\rm NL}$
\cite{jeong/komatsu:2009,*desjacques/seljak:2010}. Therefore, in the
event that {\it Planck} sees $f_{\rm NL}$ (thus ruling out single-field
models \footnote{Smidt et al. \cite{smidt/etal:2010} propose to use
$\tau_{\rm NL}=\left(\frac65f_{\rm NL}\right)^2$ as a test of
single-field inflation; however, in order to use this relation, one must
detect either $f_{\rm NL}$ or $\tau_{\rm NL}$, which would then
immediately rule out single-field models. Therefore, this relation does
not provide a test of single-field inflation. Rather, this provides a
test of inflation models where there is only a single {\it source} of
fluctuations \cite{suyama/etal:prep}, such as a curvaton scenario
as discussed also by \cite{smidt/etal:2010}.}), one of the two things can
happen: (1) $\tau_{\rm 
NL}$ is also 
detected in excess of $\frac12(\frac65f_{\rm NL})^2$, confirming
predictions from multi-field models, or (2)  $\tau_{\rm NL}$ is either 
{\it not} detected, or detected below $\frac12(\frac65f_{\rm NL})^2$,
ruling out most of the multi-field models that satisfy the above 4
conditions. This argument \cite{komatsu:2010} and our result provide a
strong science case 
for measuring the local-form trispectrum of the cosmic microwave background
as well as that of the large-scale structure of the universe.

This work is supported in part by NSF grant PHY-0758153
and by the GCOE Program ``Weaving Science Web beyond
Particle-matter Hierarchy'' at Tohoku University and
by a Grant-in-Aid for Scientific Research from JAPA 
(Nos. 18072001, 20540245 for TF) as well as by Core-to-Core Program
``International Research Network for Dark Energy.''
%\bibliography{koma_revised}
%

\end{document}